\documentstyle[12pt,sprocl,epsf]{article}
\begin{document}

{\null \vskip -1.6cm} { \null\hfill
{ORNL-CTP-95-14} {hep-ph/9509391}}

\title{ $\psi'$ and $J/\psi$ Suppression in High-Energy
Nucleon-Nucleus and Nucleus-Nucleus Collisions}

\author{ Cheuk-Yin Wong }
\address{ Oak Ridge National Laboratory, Oak Ridge, TN 37831}

\maketitle
\abstracts
{
The observed features of $\psi'$ to $J/\psi$ suppression in $pA$ and
nucleus-nucleus collisions can be explained in terms of a
two-component absorption model.  For the hard component of the
absorption due to the interaction of the produced $c\bar c$ systems
with baryons at high relative energies, the absorption cross sections
are insensitive to the radii of the $c\bar c$ systems, as described by
the Additive Quark Model.  For the soft component due to the low
energy $c \bar c$ interactions with soft particles produced by other
baryon-baryon collisions, the absorption cross sections are greater
for $\psi'$ than for $J/\psi$, because the breakup threshold for
$\psi'$ is much smaller than for $\psi$.
}

The occurrence of $J/\psi$ suppression has been suggested as a way to
probe the screening between a charm quark-antiquark pair in the
quark-gluon plasma \cite{Mat86}.  While the $J/\psi$ suppression has
been observed \cite{Bag89}, the phenomenon can be explained by
absorption models \cite{And77,Ger88,Vog91}, in which $J/\psi$-hadron
collisions lead to the breakup of the $J/\psi$ into $D \bar D X$.  A
comparison of the production of $\psi'$ with $J/\psi$ has been
suggested to distinguish between absorption and deconfinement
\cite{Gup92}.

Recent NA38 experiments \cite{Lou95,Bag95} using protons and heavy
ions at high energies reveal that $\psi'/\psi$ is
approximately a constant in $pA$ collisions \cite{Ald91}, but in SU
collisions it decreases as the transverse energy $E_T^{0}$ increases
(or equivalently, as the impact\break parameter decreases).  These features
cannot be explained by the conventional absorption models.  We would
like to describe a two-component absorption model (TCAM) which can
explain the phenomenon.

The $J/\psi$ or $\psi'$ particles are produced by the interaction of
partons of one baryon with partons of the other baryon.  The
incipient $c\bar c$ pair is created with a radial dimension of the
order of $\sim\!0.06$ fm which evolves to the bound state rms radius of
0.24 fm for $\psi$ and 0.47 fm for $\psi'$ \cite{Eic80,Bla87}.
Because $J/\psi$ is produced predominantly in the central rapidity
region, the incipient $c \bar c$ pair must have been produced
predominantly in the same rapidity region.

In soft particle production in a baryon-baryon collision, we envisage
Bjorken's inside-outside cascade picture \cite{Bjo73} or Webber's
picture of gluon branching \cite{Web84} as a quark and a diquark pull
apart after the collision, with the emission of gluons which later
hadronize.  The shape of the rapidity distribution of
produced gluons should be close to that of the produced hadrons.
Thus, produced gluons are found predominantly in the central rapidity
region.

\begin{figure}
\vskip -1.5cm
\hskip 3.5cm
\epsfxsize=250pt
\epsfbox{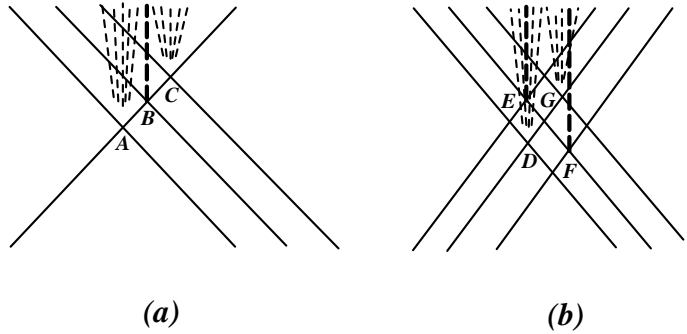}
\vskip -6.7cm
\caption{ Schematic space-time diagram in the nucleon-nucleon
center-of-mass system, with the time axis pointing upward. ($a$) is
for a $pA$ collision and ($b$) is for an $AB$ collision.  The
trajectories of the baryons are given as solid lines, the trajectories
of an incipient $c \bar c$ system produced in some of the collisions
are represented by thick dashed lines and the trajectories of soft
particles produced in some of the baryon-baryon collisions by thin
dashed lines.
\label{fig:fig1}}
\vskip -0.2cm
\end{figure}

The space-time diagram for a typical $pA$ collision is depicted
schematically in Fig.\ 1$a$ where the average trajectory of an
incipient $c \bar c$ pair does not cross the average trajectories of
soft particles produced in earlier or later collisions.  Therefore,
there is little interaction between the produced $c \bar c$ system and
these soft particles.  However, the $c \bar c$ system collides with
baryons crossing its trajectory to lead to the breakup of the $c \bar
c$ system into $D \bar DX$.  In the collision at 200A GeV, the
rapidity of a $c \bar c$ system is separated from the baryon
rapidities by about two units and the reaction cross section at these
relative energies can be calculated in the Additive Quark Model (AQM)
\cite{Lev65}.  Using the Glauber theory and a Gaussian thickness
function, the total $c\bar c$-baryon inelastic cross section in the
AQM is given by Eq.\ (12.27) of Ref.\ [16]:
\begin{eqnarray}
\label{eq:xsec}
\sigma_{\rm abs}({c\bar c{\rm -}N })
=-2\pi \beta^2
\sum_{n=1}^{6} {6 \choose n} {(- f)^n  \over n} \, ,
\end{eqnarray}
where $ f=\sigma_{cq}/\beta^2 ={\sigma_{cq} / 2\pi (\beta_{c \bar
c}^2+\beta_N^2 +\beta_{cq}^2) }\,,$ $\sigma_{cq}$ is the inelastic
cross section for the collision of $c$ (or $\bar c$) and a constituent
$q$ of the baryon, $\sqrt{3}\beta_{c \bar c}$ and $\sqrt{3} \beta_{N}$
are the rms radii of the $c \bar c$ and the baryon respectively, and
$\beta_{cq}$ is the $c$-$q$ interaction range.  For a $\psi$-$N$
absorption cross section in the range of 5 to 7 mb \cite{Ger88,Won94},
we find that $\sigma_{\rm abs}(\psi N) \sim \sigma_{\rm abs}(\psi' N)
\sim \sigma_{\rm abs}(c \bar c$-$N)$.  Thus, the absorption cross
section is approximately the same for a $c\bar c$ state during all
stages of its evolution because $6 \sigma_{cq} << 2\pi\beta^2 $.  We
ascribe the absorption due to the $(c\bar c)$-$N$ collisions at these
high relative energies as the hard component of the absorption model.
Because of the presence of only this hard component in $pA$
collisions, $\psi'$ is suppressed in the same way as $\psi$ and
$\psi'/\psi$ is a constant in $pA$ collisions , as observed
experimentally.  The approximate equality of $\sigma_{\rm abs}(\psi
N)~\sim~\sigma_{\rm abs}(\psi' N)$ at high relative energies is
further supported by the experimental ratio $\sigma_{\rm total}(\psi'
N)/\sigma_{\rm total}(\psi N)\sim 0.75$ to $0.86 \pm 0.15 $, for
$\sqrt{s}$ ranging from 6.4 GeV to 21.7 GeV \cite{Bri83,Won95b}.

To study $AB$ collisions, we adopt a row-on-row picture and consider a
typical row with a cross section of the size of the nucleon-nucleon
inelastic cross section.  The space-time diagram of the collision can
be depicted schematically in Fig.\ 1$b$.  The trajectories of the $c
\bar c$ systems cross the trajectories of colliding baryons, and the
process of absorption due to the $c \bar c$-baryon interaction (the
hard component) is the same in $pA$ as in $AB$ collisions.  However,
in $AB$ collisions, many trajectories of the produced incipient $c\bar
c$ systems cross the trajectories of the produced soft particles
(Fig.\ 1$b$).  It is necessary to consider additional interaction of
$c \bar c$ systems with soft particles in $AB$ collisions but not in
$pA$ collisions.  These interactions occur at low relative energies
and constitute the soft component of the two-component absorption
model.  At these low relative energies, the reaction thresholds
make great differences in the cross sections.  The breakup of $\psi'$,
$\chi_{1,2}$ and $J/\psi$ into $D \bar D$ requires threshold
energies of 52, $\sim\!200$, and $640$ MeV respectively.  The breakup
threshold for $\psi'$ is much smaller than for $\psi$ and
$\chi_{1,2}$.  Thus, the breakup probability due to soft particle
interactions at low energies for a $c\bar c(\psi')$ system is greater
than those for $J/\psi$ and $\chi$.  Because of this additional soft
component, in $AB$ collisions $\psi'$ is more suppressed than $\psi$,
and the suppression becomes greater as the impact
parameter decreases, as observed experimentally.

The above description of the $\psi'$ and $J/\psi$ suppression explains
the qualitative features of the suppression phenomenon.  For a
quantitative description, it is necessary to provide a description of
the produced soft particles.  It is not yet possible to ascertain the
exact nature of the soft component of the suppression mechanisms in
$AB$ collisions because of the uncertainties in the reaction cross
sections and the characteristics of produced gluons.  The soft
component of the suppression phenomenon can be attributed to (A)
produced gluons, (B) both produced gluons and hadrons, (C) produced
hadrons (as in the comover model \cite{Vog91}), or (D) deconfined
matter with no baryon absorption \cite{Kha94}.  In our model, with
$\sigma_{\rm abs}(\psi N)=\sigma_{\rm abs}(\psi' N) = 4.2$ mb fixed by
$pA$ data and a set of plausible soft particle densities and
parameters (which can be uncertain), we obtain the $\psi'/\psi$
ratio as shown in Fig.\ 2 for different scenarios with different
parameters.  Our results for the cases of (A), (B), and (C) differ by
about one percent and are represented for simplicity by a single solid
curve in Fig.\ 2.  For case (A), where we assume that the soft
particles are only gluons, the solid curve in Fig.\ 2 can be obtained
with a $\psi$-gluon absorption cross section $\sigma_{\psi g} \sim
1.4$ mb, and a $\psi'$-gluon absorption cross section $\sigma_{\psi'
g} \sim 20$ mb. For case (B) where we assume that gluons disrupt the
formation of $\psi$ and $\psi'$ while produced hadrons break up only
$\psi'$, the solid curve can be obtained with $\sigma_{\psi g} \sim
1.4$ mb, $\sigma_{\psi' g} \sim$ 7 mb and a $\psi'$-hadron
absorption cross section $\sigma_{\psi' h} \sim$ 7 mb.  For case (C),
where only produced hadrons disrupt the formation of $\psi$ and
$\psi'$, the solid curve can be obtained with $\sigma_{\psi h} \sim
0.9$ mb, and $\sigma_{\psi' h} \sim 21$ mb.  To explain the data of
$\psi'/\psi$, the parameter sets in (A)-(C) suggest greater disruption
for $\psi'$ than for $J/\psi$, due to their interaction with
soft particles.  The excessively large $\psi'$ cross sections required
to explain the $\psi'$ suppression in cases (A) and (C) may make the
(B) scenario tentatively a more attractive description.  To resolve
the ambiguities concerning gluons or hadrons, it is interesting to
note that while heavy quark production by hadron-hadron collisions is
inhibited by the OZI rule, there is no such inhibition in gluon-gluon
collisions.  The fusion of energetic gluons produced in different
baryon-baryon collisions can lead to additional charm and strangeness
production \cite{Won95} and may explain the enhanced charm and
dilepton production in $AB$ collisions relative to $pA$ collisions
observed in [21].

\begin{figure}
\vskip -0.5cm
\hskip 3.3cm
\epsfxsize=200pt
\epsfbox{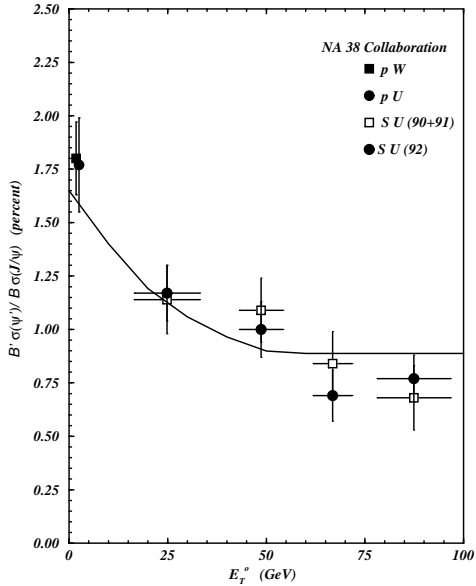}
\vskip -1cm
\caption{ The ratio ${\protect\cal B}'\sigma(\psi')/ {\protect\cal
B}\sigma(\psi)$ as a function of the transverse energy in SU
collisions at 200A GeV.  Data points are from
Ref. [7].   The theoretical results are shown as the
solid curve.
\label{fig:fig2}}
\vskip -0.4cm
\end{figure}

In conclusion, the observed features of $\psi'$ to $J/\psi$
suppression in $pA$ and $AB$ collisions can be explained in terms of a
two-component absorption model.  For the hard component of the
absorption due to the interaction of the produced $c \bar c$ systems
with nucleons at high relative energies, the absorption cross sections
are approximately the same for $\psi'$ and $J/\psi$.  However, for the
soft component of the absorption (which is due to the interaction of
the $c \bar c$ system at low relative energies with soft particles
produced by other nucleon-nucleon collisions), the absorption cross
sections are greater for $\psi'$ than for $J/\psi$ because the breakup
threshold for $\psi'$ is much smaller than for $\psi$.

\section*{Acknowledgments}

The author would like to thank T.  C. Awes, T. Barnes, and Chun Wa
Wong for helpful discussions and Dr. C. Louren\c co for valuable
comments.  This research was supported by the Division of Nuclear
Physics, U.S.D.O.E.  under Contract No. DE-AC05-84OR21400 managed by
Lockheed Martin Energy Systems.

\end{document}